\documentclass[3p]{elsarticle}
\usepackage{latexsym}
\usepackage{epsfig}
\usepackage{amssymb}
\usepackage{color}
\usepackage{graphicx} 
\usepackage{amsmath} 
\newcommand{\A}{{\mathcal A}}

\newcommand{\Lag}{{\mathcal L}}

\newcommand{\be}{\begin{equation}}
\newcommand{\ee}{\end{equation}}
\newcommand{\lsim}   {\mathrel{\mathop{\kern 0pt \rlap
  {\raise.2ex\hbox{$<$}}}
  \lower.9ex\hbox{\kern-.190em $\sim$}}}
\newcommand{\gsim}   {\mathrel{\mathop{\kern 0pt \rlap
  {\raise.2ex\hbox{$>$}}}
  \lower.9ex\hbox{\kern-.190em $\sim$}}}
\newcommand{\bw}{\begin{widetext}\begin{equation}}
\newcommand{\ew}{\end{equation}\end{widetext}}
\newcommand{\beq}{\begin{equation}}
\newcommand{\eeq}{\end{equation}}
\newcommand{\bea}{\begin{eqnarray}}
\newcommand{\eea}{\end{eqnarray}}
\newcommand{\gt}{\tilde{g}}

\newcommand{\diff}{{\rm d}}

\begin{document}
\title{Screening Vector Field Modifications of General Relativity}
\author{Jose Beltr\'an Jim\'enez$^{1,2,4}$}
\ead{jose.beltran@uclouvain.be} 

\author{ Andr\'e Lu\'is Delvas Fr\'oes$^{3,4}$}
\ead{afroes@ifi.unicamp.br}
\author{ David F. Mota$^4$}
\ead{d.f.mota@astro.uio.no}
\address{$^1$Centre for Cosmology, Particle Physics and Phenomenology, Institute of Mathematics and Physics, Louvain University $2$ Chemin du Cyclotron 1348 Louvain-la-Neuve (Belgium) \\
$^2$D\'epartement de Physique Th\'eorique and Center for Astroparticle Physics,
Universit\'e de Gen\`eve, 24 quai Ansermet, CH--1211 Gen\`eve 4
Switzerland\\
$^3$Instituto de F\'isica Gleb Wataghin, UNICAMP, 13083-859, Campinas, SP, Brazil \\
$^4$Institute for Theoretical Astrophysics, University of Oslo, P.O. Box 1029 Blindern, N-0315 Oslo, Norway}
%

%
%\date{\today}
%
\begin{abstract}
A screening mechanism for conformal vector-tensor modifications of general relativity is proposed. 
%It relies on a conformal relation between the spacetime metric and the metric to which matter couples. 
The conformal factor  depends on the norm of the vector field and makes the field to vanish in high dense regions, whereas drives it to a non-null value in low density environments. Such process occurs due to a spontaneous symmetry breaking mechanism and gives rise to both the screening of fifth forces as well as Lorentz violations.
The cosmology and local constraints are also computed.
\end{abstract}
%
%\keywords{Cosmology: Theory, Inflation, Dark Energy, N-Forms, Structure Formation}
%\pacs{98.80.-k,98.80.Jk}
%
\maketitle
%
%The quest for the correct model explaining the available cosmological observations and with satisfactory theoretical grounds is still under way. 
%The standard $\Lambda$CDM model provides excellent fits to most of the cosmological data, although some observed anomalies cannot be accommodated within it. However, it leaves theoretical cosmologists in discomfort, mainly concerning the underlying reason for the accelerated expansion that is is accounted for by simply adding a new (cosmological) constant. 
%Why would the value of the cosmological constant be so tiny as compared with the natural scale of gravity set by the Planck mass? Why would be dark energy taking over the energy density of the Universe exactly now? In an attempt to answer these questions, 
%
Driven by cosmological observations, a plethora of theoretical models have been developed in the last decades to explain the evolution and composition of the Universe.
Those models generally rely either on modifications of General Relativity (GR) or on the introduction of new exotic  components in the Universe \cite{Clifton:2011jh}.  
Often there is a mapping between the two approaches. For instance, $f(R)$ theories \cite{Sotiriou:2008rp,DeFelice:2010aj}, can be mapped via a conformal transformation into an interacting  scalar field in Einstein's  gravity. 
In this new frame, the matter fields still feel the effects of the modified gravitational interaction because the scalar field couples to them. This gives rise to a fifth force which is tightly constrained by local gravity tests.   Therefore, a general feature of novel theories to explain the nature of dark energy or dark matter, is that they modify general relativity at astrophysical scales, but are bound to recover GR at small scales via a screening mechanism. 

Several screening mechanisms have been proposed in the literature: in the {\it{chameleon}} \cite{Khoury:2003aq}, the extra scalar degree of freedom becomes more massive in regions of high densities so that its range of interaction becomes very short and the fifth force is hidden from local gravity experiments (the existence of this mechanism in $f(R)$ theories was also shown in \cite{Cembranos:2005fi}); the {\it{symmetron}} mechanism \cite{Hinterbichler:2010es} relies on an environmental-dependent potential, i.e., for low densities the potential has two minima so the field acquires a non-vanishing value, whereas for high enough densities the potential has only one minimum placed at the origin so that the scalar field vanishes; the {\it{Vainhstein mechanism}} \cite{vain} is based on kinetic self-interactions to hide the field on small scales. Recently, a mechanism to screen scalar fields via disformal couplings was proposed \cite{Koivisto:2012za}.

%That is why so much effort was put recently into developing mechanisms to evade those strong constraints, by hiding the scalar field in the Solar System, while leaving it relevant on cosmological scales. So far, three screening mechanisms like those have been developed: the chameleon\cite{KhouryC04a,KhouryC04b,MotaC07}, the symmetron\cite{HinterbichlerC10} and the Vainhstein mechanisms. The first two mechanisms resort to the introduction of an environmentally dependent potential. Thus, the chameleon mechanism provides the scalar field with a mass that depends on the surrounding density such that it becomes heavier as we go to more dense regions. The symmetron mechanism, however, introduces a potential whose shape changes with the environmental density in a similar manner to the Higgs potential. On the other hand, the Vainhstein mechanism rely on non-linear kinetic interactions to hide the field on small scales.

All the screening mechanisms in the literature have been developed for scalar degrees of freedom. Presently, there is no screening mechanism for vector-tensor modifications of GR, although chameleonic gauged B-L bosons have been discussed in \cite{Nelson:2008tn}.
However, higher spin fields are abundant in novel high energy physics theories, and have been explored in several cosmological and particle physics contexts. 
%This has been mainly motivated by the existing anomalies within the context of the $\Lambda$CDM model that seem to point out the existence of some privileged direction in the universe. 
%These observations comprise  the alignment of the low multipoles of the CMB \cite{Land:2005ad}, the hemispherical asymmetry \cite{asymmetry} or the detection of large scale bulk flows \cite{bulkflows}. 
%Among the higher spin objects, vector fields are natural candidates to try to alleviate these anomalies. 
%
In fact, they have been proposed as candidates for Lorentz violation signatures, dark energy, dark matter, inflation or as generators of curvature perturbations \cite{Clifton:2011jh, Zuntz:2008zz,Li, Bekenstein,Gies:2007su,Carroll:2004ai,Ford,vectorDE,Vectorcurvaton}. 
%However, one problem that is usually present in cosmological models with vector fields is the presence of ghost-like instabilities, although there is still room for viable models. Nevertheless, we shall not be concerned about this issue in this work since our aim is to show that it is possible to hide a vector field on small scales. Thus, 
If such fields exist, they also modify gravity and a screening mechanism is required by local gravity tests.

In this Letter, we propose  a screening mechanism for vector-tensor gravity theories, in which the vector field hides its effects on small scales while producing relevant cosmological signatures. 
%We proceed differently from \cite{NelsonVC08} in the sense that we will not use a scalar field in addition to the vector field. 
%
%In order to achieve that, we assume that the matter fields do not directly couple to the spacetime metric, but to a metric conformally related to it, being the conformal factor a function of the vector field. We show how, under certain conditions, the vector field follows a symmetron-like mechanism. 

%The model that we shall assume to develop a screening mechanism for a vector field is a vector-tensor\cite{KoivistoV08,ArmendarizV04,ArmendarizV09,,JimenezV09a,JimenezV09b,} model described by the following action:

We shall consider the simplest action to show this mechanism at work, which is  that of a massive vector field with a gauge fixing term
\be
S=\int \diff^4x\sqrt{-g}\left[-\frac{R}{16\pi G}-\frac{1}{4}F^2-\frac{1}{2}(\nabla_\mu A^\mu)^2-\frac{M^2}{2}A^2\right]+\int \diff^4x\Lag_m[\gt_{\mu\nu},\psi]\label{action},
\ee
where $R$ is the Ricci scalar constructed from the Levi-Civita connection of the metric $g_{\mu\nu}$,  $F^2=g^{\mu\alpha}g^{\nu\beta}F_{\mu\nu}F_{\alpha\beta}$ with $F_{\mu\nu}=\partial_\mu A_\nu-\partial_\nu A_\mu$ and $\Lag_m$ is the Lagrangian for the matter fields, which couple to gravity through  $\gt_{\mu\nu}$ given by\footnote{In this framework where the transformation is determined by a vector field, it is natural to consider more general couplings to matter arising from adding a disformal term such that $\tilde{g}_{\mu\nu}=B(A^2)g_{\mu\nu}+C(A^2)A_\mu A_\nu$ that could introduce novel features. Notice that, unlike for the disformal transformation involving a scalar field, no derivatives are involved in this disformal transformation.}
\be
\gt_{\mu\nu}=B^2(A^2)g_{\mu\nu},\label{confrel}
\ee
with $A^2=g^{\mu\nu}A_\mu A_\nu$. Notice that, as usual with conformal transformations, this relation guarantees that both metrics  lead to the same causal structure. In this Letter, we shall assume the particular case where the conformal factor is given by $B^2(A^2)=e^{2\beta A^2/M_p^2}$.

The action (\ref{action}) reduces to the Stueckelberg action for a massive vector field \cite{Ruegg:2003ps} in flat spacetime and when the vector field is much smaller than the Planck mass (see the Appendix). As in that scenario, one would also need to introduce the Stueckelberg field to compensate for the ghostly degree of freedom $A_0$ and, then, the theory can be quantised with a bounded hamiltonian following standard methods, in which one works with negative norm states in a restricted Hilbert space. Here, we will focus on the screening mechanism for the {\it physical} spatial components so that we have neglected the Stueckelberg field. In addition, we will show explicitly that the temporal component remains negligible in all the considered situations so that our results do not rely on its presence. Even though this does not prove the complete theoretical consistency of the full theory in an arbitrary background spacetime, we have chosen our action as a proof of concept of a working screening mechanism for a vector field.

The vector field equations of motion derived from the action (\ref{action}) can be written as
\be
\Box A_\mu=R_\mu{^\nu}A_\nu+\left(M^2+\frac{2\beta}{M_p^2}\gt^{\alpha\beta}\tilde{T}_{\alpha\beta}e^{4\beta A^2/M_p^2}\right)A_\mu\label{eomv},
\ee
where $\tilde{T}_{\alpha\beta}\equiv\frac{2}{\sqrt{-\gt}}\frac{\delta \Lag_m}{\delta\gt^{\alpha\beta}}$. This energy-momentum tensor is not conserved under the covariant derivative associated to the Levi-Civita connection generated by the spacetime metric $g_{\alpha\beta}$, i.e., $\nabla_\mu \tilde{T}^{\mu\nu}\neq0$. However, it is 
conserved under the covariant derivative associated to $\gt_{\mu\nu}$ so that $\tilde{\nabla}_\mu \tilde{T}^{\mu\nu}=0$. In other words, particles will follow the geodesics of the metric $\gt_{\mu\nu}$ and not those of $g_{\mu\nu}$. Notice that for the conformal coupling not to be trivial, it is necessary to have a dynamical norm for the vector field. Hence, this screening mechanism is not applicable to  aether theories in which the norm of the vector field is fixed by means of a Lagrange multiplier \cite{Jacobson:2000xp}.

%The $U(1)$ gauge invariance is broken by the presence of the potential and the conformal coupling. However, one can restore it by introducing a Stueckelberg field $\varphi$ and replacing $A^2\rightarrow(A_\mu-\partial_\mu\varphi)^2$ so that $A_\mu\rightarrow A_\mu+\partial_\mu\theta$ and $\varphi\rightarrow\varphi+\theta$ leaves the action invariant.
%\footnote{Due to the presence of the gauge-fixing term, the gauge parameter satisfies $\Box\theta=0$} Since the conformal coupling depends exponentially on the norm of the vector, we would have a non-local theory for the Stueckelberg field. But, the non-locality is only relevant at high energies, whereas at low energies when the exponential can be approximated by a finite number of terms of its Taylor expansion, locality is a good approximation.

Since we have two conformally related metrics, one could try to go to a {\it Jordan} frame in which matter is minimally coupled to the metric, and gravity would be described by a vector-tensor theory. In order to do that, it is necessary to invert (\ref{confrel}). However, unlike in the scalar field case, the inverted relation is not simply $g_{\mu\nu}=B^{-2}(A^2)\gt_{\mu\nu}$ because the argument of $B$ still depends on the metric $g_{\mu\nu}$. The main difficulty to invert this relation will be to solve the equation $\tilde{A}^2\equiv\gt_{\mu\nu}A^\mu A^\nu=B^2(A^2) A^2$ for $A^2$ in terms of $\tilde{A}^2$. This equation will give rise, in general, to several branches that can lead to different theories in the Jordan frame. This can be useful, for instance, to constrain the vector field to be either timelike or spacelike in such a frame without having to introduce a Lagrange multiplier and, therefore, without reducing the number of degrees of freedom. We shall not pursue the consequences of going to the Jordan frame any further here (see however \cite{EspositoFarese:2009aj}). Let us simply mention that,  in our case, the inversion of the conformal relation is given by $g_{\mu\nu}=e^{-W(2\beta \tilde{A}^2)}\gt_{\mu\nu}$, being $W(x)$ the Lambert function. 

%A crucial difference with respect to the scalar field case concerns the relation between the energy-momentum tensors associated with the two metrics: 
%\begin{eqnarray}
%T^{\mu\nu}=\frac{2}{\sqrt{-g}}\frac{\delta S}{\delta g_{\mu\nu}};\;\;\;\;\;\;\;\;
%\tilde{T}^{\mu\nu}=\frac{2}{\sqrt{-\gt}}\frac{\delta S}{\delta\gt_{\mu\nu}}.
%\end{eqnarray}

In the case of a conformal coupling depending on a scalar field $\phi$, one finds $T_{\mu\nu}=B^2(\phi)\tilde{T}_{\mu\nu}$, where $T_{\mu\nu}\equiv\frac{2}{\sqrt{-g}}\frac{\delta \Lag_m}{\delta g^{\alpha\beta}}$ is the source of Einstein's equations. However, in the case of our conformal factor depending on the vector field, we obtain
\be
T_{\mu\nu}=B^2(A^2)\left[\tilde{T}_{\mu\nu}-2B(A^2)B'(A^2)\tilde{T}A_\mu A_\nu\right],
\label{mEMT}
\ee
 where a new term arises because the conformal factor $B^2(A^2)$ depends itself on the metric. Notice that, since the additional term is proportional to the trace of  $\tilde{T}_{\mu\nu}$, it disappears for a radiation-like component. Thus, in the cosmological evolution, it can only be important when the matter component is relevant and, indeed,  it is a potential source of a large scale anisotropic stress. Nevertheless, this is not necessarily the case, since a vector field with a potential can yield an isotropic averaged energy-momentum tensor if it oscillates fast as compared to the Hubble expansion rate \cite{isotropic}

In order to understand the screening mechanism at small scales, we  consider the vector field in a Minkowski spacetime and the matter field consisting of a pressureless fluid, i.e., $\gt^{\alpha\beta}\tilde{T}_{\alpha\beta}\simeq\tilde{\rho}$. As usual with conformal couplings,  it is more convenient to use $\rho\equiv e^{3\beta A^2/M_p^2}\tilde{\rho}$, which does not depend on $A_\mu$ \cite{Khoury:2003aq,Hinterbichler:2010es}.  The field equations (\ref{eomv}) can then be written as
\be
\Box A_\mu=\left(M^2+\frac{2\beta \rho}{M_p^2} e^{\beta (A_0^2-\vec{A}^2)/M_p^2}\right)A_\mu.
\ee
We can interpret these field equations as those for a set of four scalar fields with an effective interaction that couples all four components. Even though this interaction cannot be written in terms of an effective potential as it is done in the scalar field case, the critical points can nevertheless be easily obtained
\be
A_\mu=0,\;\;\;\;\;\bar{A}^2\equiv A_0^2-\vec{A}^2=\frac{M_p^2}{\beta}\log\frac{-M^2M_p^2}{2\beta\rho}.
\label{minima}
\ee
The first critical point corresponds to a Lorentz invariant vacuum and always exists, whereas   in the second one the field acquires a non-vanishing value, thus breaking Lorentz symmetry, and only exists if $M^2$ and $\beta$ have opposite signs. Moreover, for the second critical point, we find that in regions of high density, the vector field is spacelike (timelike) for $\beta$ negative (positive).  In this Letter, we focus on the case with $\beta<0$ so that a spacelike vector field is screened and, thus, also Lorentz violations will be screened. In such a case, we can assume that $A_0\ll A_i$ so that the vector field is always space-like. This assumption is proved below to be consistent throughout the whole evolution, so that the vector field does not change from space-like to time-like. In such a case, the equations can be approximated by
\begin{eqnarray}
\Box A_\mu=\left(M^2+\frac{2\beta \rho}{M_p^2} e^{-\beta \vec{A}^2/M_p^2}\right)A_\mu.
\end{eqnarray}
Thus, we can now define the following effective potential for the spatial components
\be
V_{\rm eff}(\vec{A}^2)=-\frac{1}{2}M^2\vec{A}^2+\rho e^{-\beta \vec{A}^2/M_p^2}.
\label{Veff}
\ee
The mass matrix at the critical point with $\vec{A}=0$ is $M_{ij}=m_0^2\delta_{ij}$ with
\be
m_0^2=-M^2\left(1+\frac{2\beta\rho}{M^2M_p^2}\right).
\label{m0}
\ee
Hence, in regions of high density, the effective mass is $m_0^2\simeq-2\beta\rho/M_p^2$ (which is positive for $\beta<0$) and the VEV of the field vanishes. However, in low dense regions, we have $m_0^2\simeq-M^2<0$ and the modes become tachyonic. For the second critical point, the masses given by the eigenvalues of the mass matrix are
\be
m_1^2=m_2^2=0,\;\;\;\;\;m_3^2=2M^2\log\left[-\frac{M^2M_p^2}{2\beta\rho}\right].
\label{m3}
\ee
This result was expected since the effective potential (\ref{Veff}) exhibits an $O(3)$ symmetry that is spontaneously broken to $O(2)$ in the minimum of the effective potential, where the vector field spontaneously acquires a given value and points along one determined direction. This implies  that one massive mode and two massless modes corresponding to the unbroken symmetries arise. Since the critical point only exists for $\log\left[-\frac{M^2M_p^2}{2\beta\rho}\right]>0$, the massive mode is always non-tachyonic.
\begin{figure}[ht!]
\begin{center}
 \epsfig{width=10cm, file=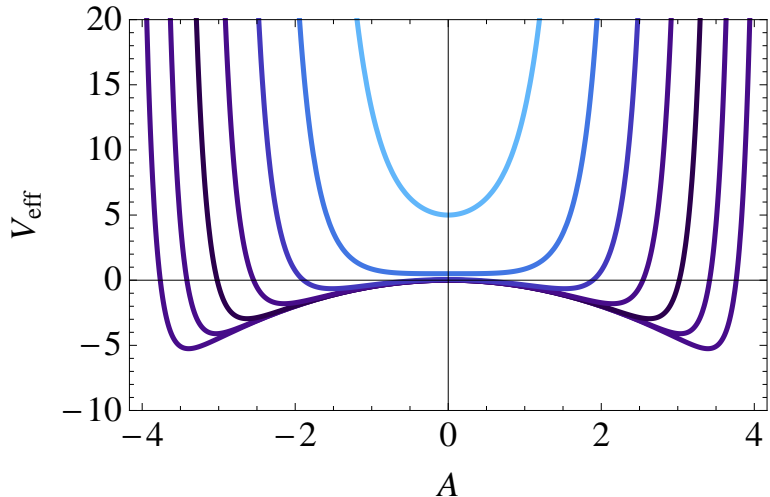}
\caption{In this plot we show the effective potential given in (\ref{Veff}) as a function of the modulus of the spatial components $|\vec{A}|\equiv A$ for different values of the energy density $\rho$. We can see how the effective potential exhibits the symmetry breaking explained in the main text when the density is low enough. Notice that at the minimum with $A\neq0$, the direction of the VEV is only spontaneously acquired so that we still have the $O(2)$ symmetry corresponding to rotations around the direction determined by the VEV of $\vec{A}$.}
\label{plotVeff}
\end{center}
\end{figure}
The effective potential thus, exhibits a symmetry breaking mechanism when going from high to low densities (see Figure \ref{plotVeff}). At very high densities ($\vert 2\beta\rho\vert\gg\vert M^2M_p^2\vert$), the effective potential has only one critical point located at the origin so that the vector field has a vanishing VEV and a positive mass $m_0^2\simeq\frac{2\vert\beta\vert\rho}{M^2M_p^2}$.  On the other hand, when the density is low enough ($\vert 2\beta\rho\vert\ll\vert M^2M_p^2\vert$), the critical point at the origin becomes tachyonic and the field runs away from it. However, the conformal coupling stabilises this tachyonic evolution and  another set of critical points appear. In the new vacuum corresponding to the effective potential after the symmetry breaking, we have two massless modes plus a massive mode with mass $m_3^2$ given in (\ref{m3}). 

The difference of our screening mechanism with respect to the symmetron is that, in the symmetron, the stabilisation of the tachyonic mode after the symmetry breaking is done by the $\phi^4$-term of the potential, whereas in our case the stabilisation comes from the exponential term of the conformal coupling. The fact that the stabilisation is done by an exponential instead of a quartic term is the reason why our screening process is more efficient after the symmetry breaking than in the original symmetron, although this could be very straightforwardly adapted for the symmetron.

The aforementioned critical points for the effective potential are the key for the screening mechanism: provided that $-M^2M_p^2\ll{2\beta\rho_{\rm local}}$ for the local density in the Solar System or our galaxy, the VEV of the vector field is zero. The leading order of the interaction of the vector field with the matter fields is $\beta A_{\rm VEV}^\mu\delta A_\mu T/M_p^2$  so that it decouples from matter in high density environments. When the vector field acquires a non-vanishing VEV, the strength of the interaction is set by $\beta A_{\rm VEV}^\mu/M_p^2$ so if we want this interaction to be of the same order as that of gravity, we need, at least, $\beta A_{\rm VEV}^\mu\sim M_p$. This condition is indeed fulfilled when the symmetry is broken and the field acquires the VEV given in (\ref{minima}), since the condition for the symmetry breaking is precisely $-\frac{M^2M_p^2}{2\beta\rho}\gsim1$. It is also interesting to note that due to the conformal invariance of electromagnetism in 4 dimensions, photons do not couple to the vector field because of the tracelessness of its energy-momentum tensor. Finally, we should stress the fact that the interaction is direction dependent and only the component of the vector field parallel to its VEV couples to matter. 

Another interesting feature of this mechanism is that, after the symmetry breaking takes place, we have a non-vanishing VEV for the vector field, so there is also a spontaneous breaking of Lorentz symmetry\footnote{The breaking of Lorentz symmetry to which we refer in this Letter and that is common in the literature actually refers to a breaking of isotropy in the vacuum.} when going from high to low density regions. This represents a distinctive feature with respect to screening mechanisms for scalar fields. In fact,  this mechanism can be regarded not only as a way to screen the fifth force mediated by the vector field, but also as a mechanism to screen Lorentz violations or, in other words, as a mechanism to dynamically restore Lorentz invariance in high density regions, while being broken in low density environments. Screening of Lorentz violating interactions has also been explored in \cite{Brax:2012hm} in the context of modified gravity theories with a scalar field $\phi$. In that case, the Lorentz violating coupling is through a coupling term $\partial_\mu\phi \partial_\nu\phi T^{\mu\nu}$ so that a direction dependent interaction appears when there is a background with non-vanishing gradients of the scalar field.

The local  bounds on the theory can be computed by calculating the field profile near a static and spherically symmetric object. The field equations read
\begin{eqnarray}
A_0''+\frac{2}{r}A_0'=-\left[M^2+2\frac{\rho\beta}{M_p^2} e^{\beta A^2/M_p^2}\right]A_0\\
A_z''+\frac{2}{r}A_z'=-\left[M^2+2\frac{\rho\beta}{M_p^2} e^{\beta A^2/M_p^2}\right]A_z.
\end{eqnarray}
To obtain the profile, we shall solve these equations outside and inside a spherical object. In the outer region, we assume that we are in the phase of symmetry breaking so that we will expand the equations to linearize them around the corresponding critical point. In such a case, the equations for the perturbations of the field with respect to their values at the fixed point are
\begin{eqnarray}
\delta A_0''+\frac{2}{r}\delta A_0'=\frac{2\beta M^2\bar{A}_0}{M_p^2}\left(\bar{A}_0\delta A_0-\bar{A}_z\delta A_z\right)\\
\delta A_z''+\frac{2}{r}\delta A_z'=\frac{2\beta M^2\bar{A}_z}{M_p^2}\left(\bar{A}_0\delta A_0-\bar{A}_z\delta A_z\right)
\end{eqnarray}
where $\delta A_0\equiv A_0-\bar{A}_0$, $\delta A_z\equiv A_z-\bar{A}_z$ and $\bar{A}_0$, $\bar{A}_z$ are the asymptotic values. Now, as we commented above, we can assume that $\bar{A}_0\ll\bar{A_z}$ because the vector field is spacelike at the critical point. Then, assuming that the perturbations on both components are of the same order, we can further approximate the above equations to obtain
\begin{eqnarray}
&&\delta A_0''+\frac{2}{r}\delta A_0'\simeq m_3^2\frac{\bar{A}_0}{\bar{A}_z}\delta A_z,\\
&&\delta A_z''+\frac{2}{r}\delta A_z'\simeq m_3^2\delta A_z.
\end{eqnarray}
From these equations we can see that the source term for $\delta A_0$ is determined by $\delta A_z$ so that its profile will follow that of $\delta A_z$. Thus, since its asymptotic (cosmological) value $\bar{A}_0$ is assumed to be smaller than $\bar{A}_z$, it will remain smaller for all $r$. It is important to notice that both components satisfy the same equation inside the object, so that this remains true also in the inner region. 

\begin{figure}[ht!]
\begin{center}
 \epsfig{width=10cm, file=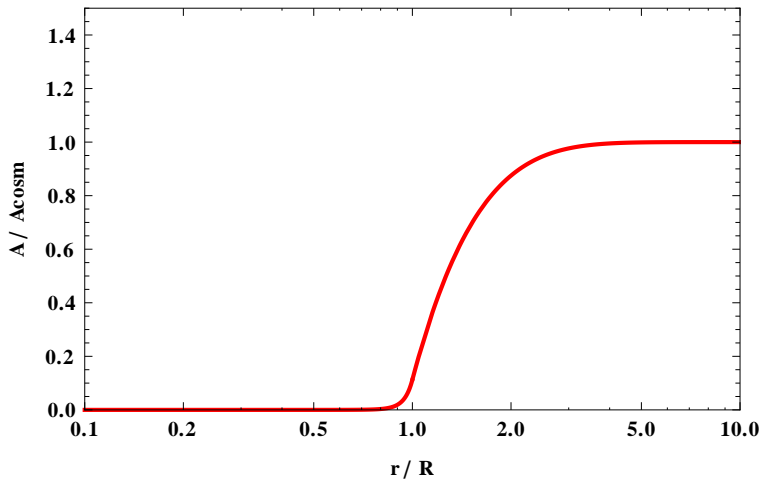}
\caption{The field profile inside and outside the galaxy: the field goes to zero inside the galaxy whereas it tends to a non-vanishing (cosmological) value outside. Thus, inside the galaxy, the fifth force is screened and Lorentz symmetry is restored. As explained in the main text, the temporal component profile is driven by the spatial component so that if it is asymptotically smaller on cosmological scales, it will remain small at all scales and both outside and inside the object.}
\label{profile}
\end{center}
\end{figure}

The obtained equations look similar to those in \cite{Hinterbichler:2010es} so we can proceed in a similar manner to obtain the solutions inside and outside the object. The corresponding solutions for an object of density $\rho_R$ and size $R$ and with the boundary conditions $A'_z(0)=0$ and $A(r\rightarrow\infty)\rightarrow\tilde{A}$, being $\tilde{A}$ the cosmological value, can be written as follows:
\begin{eqnarray}
&&A_z^{\rm in}=B\frac{R}{r}\sinh\left[\frac{r}{R}\sqrt{\alpha^2-\mu^2}\right], \\
&&A_z^{\rm out}=C\frac{R}{r}e^{-m_3r}+\bar{A}_z
\nonumber
\end{eqnarray}
where $\alpha^2\equiv-2\beta\rho_R R^2/M_p^2$ and $\mu^2\equiv M^2R^2$  are dimensionless parameters. The mass $m_3$ is the one given in (\ref{m3}). It is important to notice that in the expression for $m_3$ we need to use the cosmological density and not  $\rho_R$. On the other hand, we focus on the massive mode because the massless modes are less constraining. Finally, the constants $B$ and $C$ are obtained so that both solutions (and their first derivatives) match at $r=R$. In Figure \ref{profile} we show the profile corresponding to the above solution where we see the analogue of the thin-shell effect and how the field profile goes to zero very rapidly inside the object.

The force acting on a test particle due to the vector field is given by the gradient of the conformal factor $\vert \vec{F}\vert=\diff \ln B/\diff r$, as obtained from the geodesic equations.  We should notice here that, unlike in the scalar field case, the full expression for the fifth force actually depends on the gravitational potential because $B(A^2)$ depends on the metric. If we assume the weak field limit with $g_{\mu\nu}=\eta_{\mu\nu}+h_{\mu\nu}$ where $\vert h_{\mu\nu}\vert\ll1$, then the fifth force is given by
\begin{eqnarray}
\vert\vec{F}\vert&=&\frac{\beta}{M_p^2}\left[2\Big(\eta^{\mu\nu}-h^{\mu\nu}\Big)A_\mu \frac{\diff A_\nu}{\diff r}-\frac{\diff h^{\mu\nu}}{\diff r}A_\mu A_\nu\right]\\\nonumber
&\simeq&\frac{2\beta}{M_p^2}\left[A_0 \frac{\diff A_0}{\diff r}-\vec{A}\cdot\frac{\diff\vec{A}}{\diff r}-\frac 12\frac{\diff h^{\mu\nu}}{\diff r}A_\mu A_\nu\right].
\end{eqnarray}
The last term in the bracket is proportional to the usual gravitational force given by the gradient of the gravitational potential so that it can be seen as a modification of Newton's constant. Thus, the leading term of the fifth force gives a term proportional to the value of the field and its gradient and it also modifies the effective Newton's constant. However, the effects are negligible inside the galaxy, since there the value of the field drops dramatically and we have a thin-shell effect (see Figure \ref{profile}).

\begin{figure}[ht]
\begin{center}
\epsfig{width=8cm, file=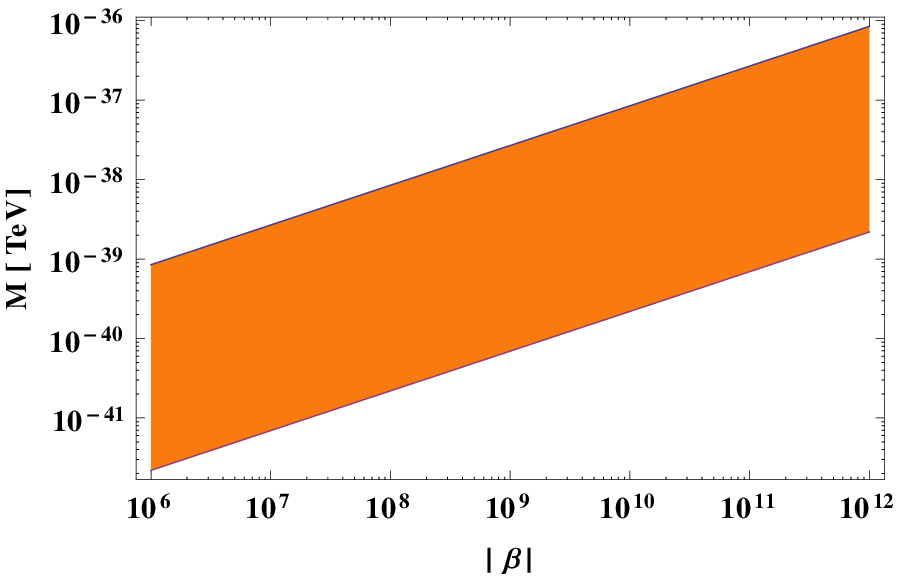}
\epsfig{width=8cm, file=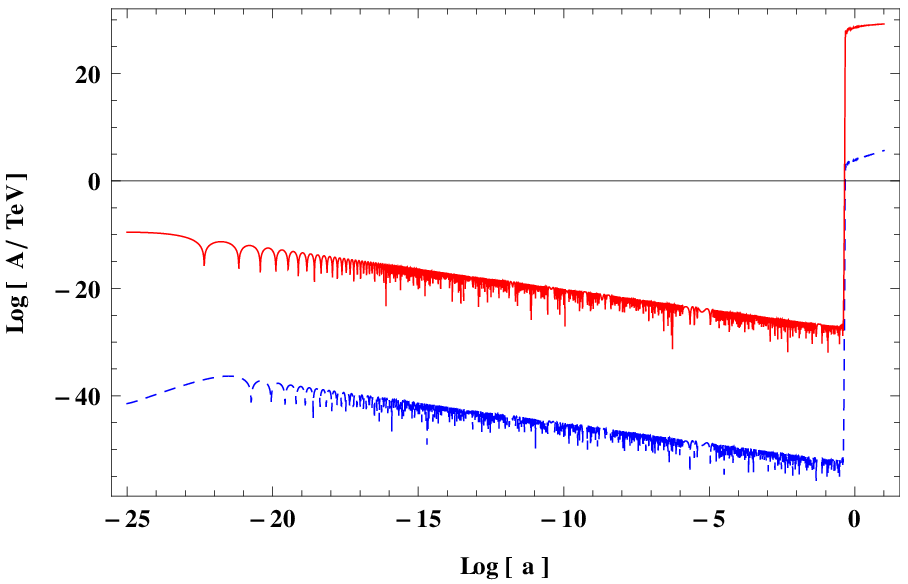}
\caption{Left panel: Parameter space, for   $|\beta|>10^6$ in which the symmetry is  broken outside the galaxy and unbroken inside so that fifth force and Lorentz violations are screened. Right panel:  Cosmological evolution of the temporal (dashed blue) and the spatial (solid red) components of the vector field, where we can see how the symmetry breaking takes place and the field goes to the new minimum and oscillates there.}
\label{parameterspace}
\end{center}
\end{figure}

Our screening  process drives the fifth force due to the vector field to extremely small values, away from any possible detectability range in present or near future experiments. 
%For that we only impose $|\beta|>10^5$, which is the limit for which the vector is sufficiently screened inside the galaxy. 
As an extreme illustration, in  Figure \ref{parameterspace},  we used $|\beta|>10^6$ and $M\gsim H_0\sim 10^{-42}$ TeV. It is clear from the figure that weak equivalence principle violations and E\"{o}t-Wash like experiments are easily satisfied for a wide range of the theory parameter space. 
%Any bigger $|\beta|$ will only make the mechanism more efficient. 
For $|\beta|=10^6$ the value of the field is $\sim10^{-7}$ TeV inside the galaxy, and the profile is almost flat, so the gradient of the field is negligible. Inside high density objects like the Sun, the thin shell effect is even more efficient.
 %

%\vspace{-1cm}

%ConcerningBig Bang nucleosynthesis constraints is also satisfied as any possible variation of the mass of the particles would vary proportional to $B(A^2)$, which is very small inside high density environments.

%The field profile obtained this way can be seen in figure \ref{profile} below. We see that the field value falls in a thin-shell manner, and we observe that a bigger $|\beta|$ means the field minimum inside the galaxy goes to values nearer zero.

%Which is proportional to the field value times it's derivative in space. As the field value goes to the minimum (zero) inside the galaxy with a thin shell, this means we have a tiny, undetectable fifth force inside the galaxy.

Notice that, although mediated by a vector field, the test particle feels a force depending only on the magnitude of the vector field, but not on its direction. This is so because,  the coupling to matter is through $A^2$ to the trace of the energy-momentum tensor. However, the vector nature of the new interaction appears by means of gravitational effects, i.e., when the backreaction of the vector field on the metric is relevant and we have a spontaneous breaking of Lorentz invariance,  on cosmological scales when the energy density drops below the critical value leading to the symmetry breaking phase.

Although the field hides in high density regions, at cosmological scales the symmetry can be broken and astrophysical signatures appear. In the early universe (radiation and matter dominated epochs), we expect the vector field to be subdominant. In such a case, it is justified to consider the equations for the vector field and assume that the expansion is driven by the dominant component. As before, we assume that the vector field is spacelike. This means that we have $\A_i\equiv A_i/a\gg A_0$. Moreover, we  assume that the spatial component has linear polarisation along the $z$-direction. Then, in a FLRW metric, the equations of motion are given by\footnote{Here we introduce $T=e^{3\beta A^2/M_{p}}\tilde{T}$ as usual.}
\begin{eqnarray}
\ddot{A}_{0}+3H\dot{A}_{0}&=&\left[-3\dot{H}+M^2+2\frac{T\beta}{M_{pl}^2}e^{\beta(A_0^2- \A_z^2)/M_{p}^2}\right]A_{0}\\
\ddot{\A}_z+3H\dot{\A}_z&=&\left[M^2-\dot{H}-2H^2+2\frac{T\beta}{M_{p}^2}e^{\beta(A_0^2- \A_z^2)/M_{p}^2}\right]\A_z.\nonumber
\end{eqnarray} 
We have solved these equations numerically throughout the universe evolution and the solutions are plotted in Figure \ref{parameterspace}. In the following we shall find approximate analytical solutions for the cosmological evolution. Since the coupling to the background density is through the trace of the energy-momentum tensor, only non-relativistic species are relevant for the interactions of the vector field, i.e., only the matter component will modify the effective mass of the vector field. Thus, we can simply set $T=\rho_m$, even in the radiation dominated epoch. Moreover, we also assume that $A_0^2, \A_z^2\ll M_p^2/\beta$ so that we can approximate the exponential by 1. Notice that, as long as this condition holds, the equations decouple and both components evolve independently. Well inside the radiation dominated epoch the equations can be approximated by
 \begin{eqnarray}
\ddot{A}_{0}+3H\dot{A}_{0}&\simeq&-3\dot{H}A_{0},\\
\ddot{\A}_z+3H\dot{\A}_z&\simeq&0\nonumber
\end{eqnarray}
where we have used that $T\simeq0$ and $H^2\gg M^2$. Thus, the growing mode for the temporal component is given by $A_0\propto t$, whereas $\A_z$ remains frozen. This behaviour goes on until the energy density of matter becomes relevant in the field equations. This will imply that we need to tune the initial conditions such that, even though $A_0$ grows with respect to $\A_z$, it remains smaller ($A_0<\A_z$) by the time when $\rho_m$ starts being relevant. If we look at the equations, we can see that this happens when $2\beta\rho_m/M_p^2\simeq H^2\simeq\rho_r/M_p^2$, i.e., when $\rho_r\simeq2\beta\rho_m$. The corresponding redshift is
$(1+z_T)=2\beta\Omega_m/\Omega_r=2\beta(1+z_{\rm eq})$ with $z_{\rm eq}$  the equality redshift. Thus, provided that $\beta>1$, this indeed happens in the radiation dominated epoch. After this time, the matter component becomes relevant and the equations can be approximated by:
\begin{eqnarray}
\ddot{A}_{0}+3H\dot{A}_{0}&=&\frac{2\beta \rho_m}{M_{pl}^2}A_{0},\\
\ddot{\A}_z+3H\dot{\A}_z&=&\frac{(2\beta-\frac{1}{6}) \rho_m}{M_{pl}^2}\A_z,
\end{eqnarray}
where we have used that $\dot{H}+2H^2=\frac{1}{6M_p^2}T$. Since we are assuming a negative value for $\beta$, the solutions are damped oscillations for both components. During radiation, the damping factor is $t^{-1/4}$, whereas in the matter epoch it is $t^{-1/2}$ for both components. The field behaves in this way until $M^2\sim \beta H^2$. At this point, the effective mass becomes negative and the field grows exponentially until $\A^2\sim \vert M_p^2/\beta\vert$, when the higher order terms become important. This happens because the critical density at which the symmetry breaking occurs is reached and the field evolves towards the new minimum. Once the new minimum is reached, the field starts oscillating around it. However, since the position of the minimum is time-dependent, with a timescale of order $\beta \rho/M_p^2$, the center of the oscillations moves. This is so because the oscillations timescale is 
$$\tau^{-2}\sim m_{\rm eff}^2\sim M^2\log\frac{-M^2M_p^2}{2\beta\rho}\sim M^2\log\frac{-M^2}{2\beta H}$$ 
and the timescale associated with the evolution of the minimum is $H^{-1}$, so that we have that $\frac{m_{\rm eff}^2}{H^2}\sim  \frac{M^2}{H^2}\log\frac{-M^2}{2\beta H^2}$, that is large after the symmetry breaking. Therefore, the minimum evolves adiabatically with respect to the field oscillations.
%
%\begin{figure}[hb!]
%\begin{center}
 %\epsfig{width=7cm, file=LogA0Az.eps}
%\caption{Cosmological evolution of the vector field components.The initial conditions are such $\A_i\gg A_0$ (this remains valid throughout the whole evolution).  We choose a linear polarisation along the z-axis, therefore, the two lines correspond to $A_0$ and $A_z$. When the background density becomes low enough, symmetry breaking takes place.Then the field goes to the new minimum of the potential where it oscillates.}
%\label{cosm}
%\end{center}
%\end{figure}

Notice that during Big Bang nucleosynthesis the field is cosmologically screened so that no effects are present and the corresponding constraints are easily evaded. 
%constraints is also satisfied as any possible variation of the mass of the particles would vary proportional to $B(A^2)$, which is very small inside high density environments.

%

%These theories leave however a signature at astrophysical scales. This can be seen by taking the trace of the Einstein equations, in the case where the timelike component of the vector field is subdominant:
%\be
%\nonumber
%R=\frac{1}{M_{pl}^{2}}\left(e^{\frac{-2\beta A_{z}^{2}}{a^{2} M_{pl}^{2}}}+\frac{2 \beta A_{z}^{2}}{a^2}e^{\frac{-4\beta A_{z}^{2}}{a^{2}M_{pl}^{2}}}\right)T+2M^{2}A_{z}^{2}
%\ee

%The expression above is just the new energy momentum tensor defined above (\ref{mEMT}), plus a potential like term dependent on the spacelike component of the vector field. Comparing with the usual expression for this trace in $\Lambda CDM$, $R = G \ T$, we observe that when the field is subdominant cosmologically or well inside the galaxy, we have the same behaviour. However, after symmetry breaking, with the value of the field increasing, the relation is modified.

%We can therefore define a $G_{eff}$
%
%\be
%\nonumber
%G_{eff}=\frac{1}{M_{pl}^{2}}\left(e^{-2\beta A_{z}^{2}/a^{2}M_{pl}^{2}}+\frac{2 \beta A_{z}^{2}}{a^{2}}e^{-4\beta A_{z}^{2}/a^{2}M_{pl}^{2}}\right)
%\ee

The main difference between these models and GR comes from an anisotropic effective gravitational constant which will affect structure formation at large scales. Moreover, these models will have imprints in cosmology which are not present in other screening models, such as the chameleons and symmetron type of models. Those signatures will arise from the novel extra-term in the energy-momentum tensor (\ref{mEMT}) proportional to $A_{\mu}A_{\nu}$ that can produce anisotropic stresses on large scales that will contribute to the ISW effect.

In summary, a screening mechanism for conformal vector-tensor modifications of general relativity is proposed.  Such mechanism allows to screen a vector field on small scales while non-trivial cosmological effects can still be present due to modifications of Einstein's equations.  We focus on a simple model consisting of a massive vector field which is conformally coupled to matter. The screening mechanism occurs due to a spontaneous symmetry breaking, therefore is applicable to a whole class of theories with different combinations of potential and conformal couplings so that the vector field is either timelike or spacelike at the critical points after the symmetry breaking. Moreover, our mechanisms also provides a way to restore Lorentz invariance in high dense regions, while  being broken in low dense regions. This is a novel and unique signature of this mechanism.
%Furthermore, general couplings to matter could be introduced, like adding a disformal term such that $\tilde{g}_{\mu\nu}=B(A^2)g_{\mu\nu}+C(A^2)A_\mu A_\nu$ that could introduce novel features. 
Notice also that in spite of local gravity constraints being easily evaded, the cosmological structure formation within these theories will be different from both General Relativity and other screened-modified gravities, due to the coupling between matter and the vector field.
%: the main signatures come from the novel extra-term in the energy-momentum tensor and an effective gravitational constant which lead to modifications in cosmological structure formation. 
Finally the screening mechanism operating
on small scales opens a new avenue for fundamental vector fields strongly coupled to matter into our theories.

%The screening mechanism proposed here for a vector field is somehow a mixture of the chameleon and symmetron models for scalar fields.

\section*{Acknowledgments}
JBJ is supported by Wallonia-Brussels Federation grant ARC No.~11/15-040. JBJ also thanks  the Spanish MICINNÕs Consolider-Ingenio 2010 Programme MultiDark CSD2009-00064 and project number FIS2011-23000 and wishes to thank the Instituto de F\'isica Gleb Wataghin for their hospitality. ALDF is supported by CAPES agency, through the grant BEX 0070/10-6.
D.F.M. thanks the Research Council of Norway FRINAT grant 197251/V30.
D.F.M. is also partially supported by project CERN/FP/123618/2011 and CERN/FP/123615/2011.
%DFM is also partially supported by the projects PTDC/FIS/111725/2009 and CERN/FP/116398/2010.
%JBJ is supported by the S 
%contract EX2009-0305
 
% Programme MultiDark CSD2009-00064. 
%
%\vspace{-1cm}

\section*{Appendix}
In this  Appendix we intend to briefly comment on the relation of our action with the Stueckelberg action for a massive vector field. To show that, we consider the theory in flat spacetime and that the vector field is much smaller than the Planck mass. In such a limit, the conformal relation can be approximated by
\be
\gt_{\mu\nu}\simeq\left(1+\frac{2\beta A^2}{M_p^2}\right)\eta_{\mu\nu},
\ee
with $\eta_{\mu\nu}$ the Minkowski metric. Then, we can plug this expression into the action to obtain the following action for the vector field
\be
S=\int d^4x\left[-\frac{1}{4}F^2-\frac{1}{2}(\partial_\mu A^\mu)^2+\frac 12 M_{\rm eff}A^2]\right].
\ee
where we have defined
\be
M_{\rm eff}^2\equiv-\frac{M^2}{2}\left(1+\frac{2\beta}{M_p^2}\gt^{\alpha\beta}\tilde{T}_{\alpha\beta}\right).
\ee
Notice that this is nothing but the mass $m_0^2$ obtained for the field in the phase without symmetry breaking.  Now, we remind the action for a massive vector field in the Stueckelberg formalism (see for instance  \cite{Ruegg:2003ps})
\be
S=\frac12\int d^4x\left[-\partial_\mu A_\nu\partial^\mu A^\nu +m^2A^2+\partial_\mu \varphi\partial^\mu \varphi-m^2 \varphi^2\right]
\label{action2}
\ee 
where $\varphi$ is a scalar field (the Stueckelberg field). After integrating by parts, dropping surface terms and recasting the resulting terms, the Stueckelberg action can be written as
\be
S=\int d^4x\left[-\frac{1}{4}F^2-\frac{1}{2}\left(\partial_\mu A^\mu+m\varphi\right)^2+\frac{m^2}{2}\left(A_\mu+\frac 1m\partial_\mu \varphi\right)^2\right].
\label{action3}
\ee
From this form of the action, it is more apparent that it exhibits a gauge symmetry $A_\mu\rightarrow A_\mu+\partial_\mu\Lambda$, $\varphi\rightarrow \varphi+m\Lambda$, with the gauge function satisfying the wave equation $(\Box+m^2)\Lambda=0$. It is also apparent that our action reduces to Stueckelberg action in a gauge with $\varphi=0$, which can be chosen since the restriction of the gauge parameter coincides with the equation of $\varphi$. Another way of interpreting our action is as the Stueckelberg action in which we neglect $\varphi$ (not necessarily imposed by a gauge condition). We also remind here that to guarantee the consistency of the theory at the quantum level, it is necessary to impose the following additional subsidiary condition:
\be
(\partial_\mu A^\mu+m\varphi)^{(-)}\big\vert{\mathrm{ phys}}\big\rangle=0
\ee
where,  the superscript $^{(-)}$ denotes the positive frequency part of the operator (i.e., it only involves annihilation operators) and $\big\vert{\mathrm{ phys}}\big\rangle$  is the space of physical states. This is nothing but the analogous of the Gupta-Bleuler condition implemented for the massive case. Thus, one works in a space of indefinite metric, but the physical states have positive norm and the Hamiltonian is also positive definite on the physical space.

 Instead of using our action without the Stueckelberg field, one could alternatively consider the full action including $\varphi$. In such a case, to maintain the gauge invariance, the conformal relation should also include the Stueckelberg field
\be
\gt_{\mu\nu}=B^2(x)g_{\mu\nu},
\ee
where the argument of the conformal factor is $x\equiv (A_\mu+\frac 1m\partial_\mu \varphi)^2$.

It is important to notice that going from the flat spacetime version of the action for a massive vector field to its curved spacetime version is not free from the usual ambiguity when covariantising a given action by replacing ordinary partial derivatives with covariant derivatives. One could covariantise either (\ref{action2}) or (\ref{action3}) and one would end up with different curved spacetime versions of the same  theory in flat spacetime. The difference between both actions will be a non-minimal coupling of the vector field to the curvature, more precisely, a term $R_{\mu\nu}A^\mu A^\nu$. Although this term would give rise to a different cosmological evolution or different features in contexts where curvature is relevant, it is important to note that,  since it vanishes in a Minkowski spacetime, it does not affect the screening mechanism proposed in this Letter, which would equally work for both versions of the covariant action.

\section*{References}
%\bibliography{Vector}
%

\end{document}